\documentclass[prl,a4paper,twocolumn,showpacs,superscriptaddress,floatfix]{revtex4}
\usepackage{times}

\usepackage{graphicx}
\usepackage{amsmath}
\usepackage{amssymb}
\usepackage{bm}
\newcommand{\be}{\begin{equation}}
\newcommand{\ee}{\end{equation}}

\begin{document}

\title{Anti-correlation for Conductance Fluctuations in Chaotic Quantum Dots}

\author{A. L. R. Barbosa}
\affiliation{Departamento de F\'sica, Universidade Federal Rural de Pernambuco, Dois Irm\~aos, 52171-900, Recife, PE, Brazil}

\author{M. S. Hussein}
\affiliation{Instituto de Estudos Avan\c cados and Instituto de F\'{\i}sica, Universidade de S\~{a}o Paulo,
  C.P.\ 66318, 05314-970 S\~{a}o Paulo, SP, Brazil}

\author{J. G. G. S. Ramos}
\affiliation{Departamento de Ci\^encias Exatas, Universidade Federal da Para\'{\i}ba, 58297-000, Rio Tinto, PB, Brazil}

\date{\today}

\begin{abstract}

We investigate the correlation functions of mesoscopic electronic transport in open chaotic quantum dots with finite tunnel barriers in the crossover between Wigner-Dyson ensembles. Using an analytical stub formalism, we show the emergence of a depletion/amplification of conductance fluctuations as a function of tunnel barriers, for both parametric variations of electron energy or magnetoconductance fields. Furthermore, even for pure Dyson ensembles, correlation functions of conductance fluctuations in chaotic quantum dots can exhibit anti-correlation. Experimental support to our findings is pointed out.

\end{abstract}


\maketitle

{\it Introduction} - Statistical properties of open quantum systems has been widely studied over recent decades \cite{Weidenmuller}. The aleatory spectrum of the chaotic resonant scattering in heavy nuclei \cite{Bohr} led Wigner to discover the random nature of Hamiltonians \cite{Richter1}. More recently, the quantum chaotic scattering becomes a very precise investigation subject through experimental control of artificial mesoscopic Quantum Dots (QD) \cite{Heinzel,Weidenmuller}.

The coherent transport in QDs has many subtle interference patterns effects resulting from multiply quantum scattered waves, and the universal conductance fluctuations (UCF) is one of the most common \cite{Alhassid}. The UCFs in open ballistic QDs are random values arising when a external parameter, such as a magnetic field ${\cal B}$ or the electronic energy $\varepsilon$, is varied. It was found that these sample-to-sample random fluctuations in conductance $g(\varepsilon,{\cal B})$ depend solely on fundamental symmetries of the nature, which were classified by Wigner-Dyson in universal orthogonal ($\beta=1$), unitary ($\beta=2$) and symplectic ($\beta=4$) ensembles \cite{Weidenmuller,Beenakker}.

Random Matrix Theory (RMT) \cite{Weidenmuller, Richter1} within the Landauer-B\"uttiker framework \cite{Buttiker,Beenakker} can explain an observed UCF and its characteristic correlation function (CF). The current theoretical endeavors infers that CF amplitude (variance) of an UCF depend uniquely, and in an essential way, from the Dyson symmetry index $\beta$. Also infers that their shapes in Dyson' ensembles can be Lorentzian \cite{Blumel,Baranger}, $C(\delta \varepsilon) \propto (1+\delta \varepsilon^2)^{-1}$ for CF of electronic energy UCFs, or square-Lorentzian \cite{Baranger, efetov}, $C(\delta {\cal B}) \propto (1+\delta {\cal B}^2)^{-2}$ for CF of magnetoconductance UCFs.
\begin{figure}[h!]
	\centering
	        \vskip-0.4cm
		\includegraphics[width=\columnwidth,height=5.0cm]{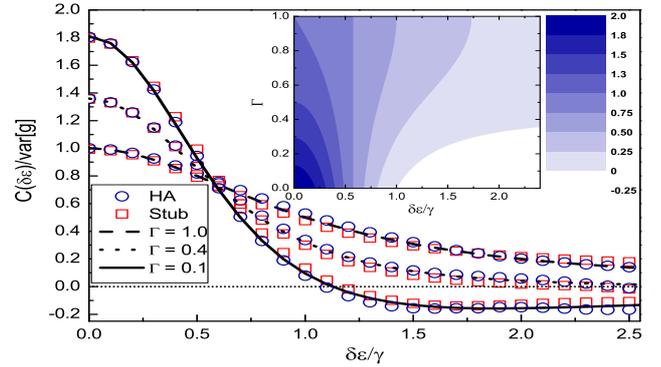}
	\caption{(Color online) Correlation function as a function of parametric variation of energy. Transition from Lorentzian (long tracejated line) to a anti-correlation (continuous line) as a function of symmetric $\Gamma$. The inset diagram $\delta \varepsilon/\gamma \times \Gamma$ show values of correlation function in each color.}
	\label{Fig1}
\end{figure}

The shape of the Lorentzian has implications on the density of maxima of an UCF trace \cite{gabrielprl}, acting as signatures of quantum chaos that emerges from randomness and long-range repulsion of energy levels of a QD random Hamiltonian. This behavior, common for each Dyson index $\beta$, invites one to ask if there would be some kind of simple quantum parameter which strikingly modify the lorentzian-shape and the amplitudes of CFs. As we analytically show, the tunnel barriers \cite{Marcus,Cronenwett,Gustin,Crook,Zumbuhl} performs a fundamental role in this question.

In open chaotic balistic QDs with irregular borders, the tunneling barriers characterizes the degree of openness of the QD, determined by the coupling between the resonant modes of the QD and the open channels of the connected leads. Their effects can be encoded on tunneling probabilities $\Gamma$ of electronic entrance on the QD \cite{Marcus,Zumbuhl,Gustavsson}. As standard, we denote the tunnel probability by $\Gamma \in [ 0,1 ]$, which assumes asymptotic value $\Gamma \rightarrow 1$ for ideal contacts, and $\Gamma \rightarrow 0$ for QD with high reflection probabilities.

An example of our results is the following quite simple formula for the normalized CF in the ``quasi-closed" QD \cite{Zirnbauer}, obtained in the limit that the number of open channels $N_{i}$ is large and tunneling probabilities $\Gamma_{i}$ of the lead $i=1,2$ tend to zero,
\begin{eqnarray}
\frac{C(\delta \varepsilon)}{var[g]} &=& \frac{1-(\delta \varepsilon/\gamma)^2}{\left[1+(\delta \varepsilon/\gamma)^2\right]^2} \label{covg}
\end{eqnarray}
where $var[g]=4/\beta \cdot (G_1G_2)^2/ (G_1+G_2)^4$ is the variance of the UCFs and $G_i=N_i\Gamma_i$. There is a simple expression that relates the auto-correlation length $\gamma$ to the mean resonance spacing $\Delta$, namely $\gamma= (G_1+G_2)\Delta/2\pi$. If $\delta \varepsilon/\gamma> 1$, notice that the correlation is negative indicating there is an anti-correlation in the UCF.

The strongly non-Lorentzian shape of a conductance correlation function is clearly seen in Fig.[\ref{Fig1}] for some values of the tunneling probabilities. This behavior is further supported by analytical and numerical results described below.

{\it Integration over the unitary group} - We consider two-dimensional QDs in the presence of both full spin rotation and time reversal \cite{TReversal,Schafer} symmetries (${\cal B}=0$, $\beta=1$) or in time reversal completely broken (${\cal B} \rightarrow \infty$, $\beta=2$), connected to electron reservoirs via two ideal leads. We also consider QDs in the presence of ${\cal B}$ finite, giving rise to a crossover \cite{FS2} between the Wigner-Dyson pure ensembles.

\begin{figure}[!]
	\centering
	        \vskip-0.4cm
           \includegraphics[width=7.0cm,height=0.8cm]{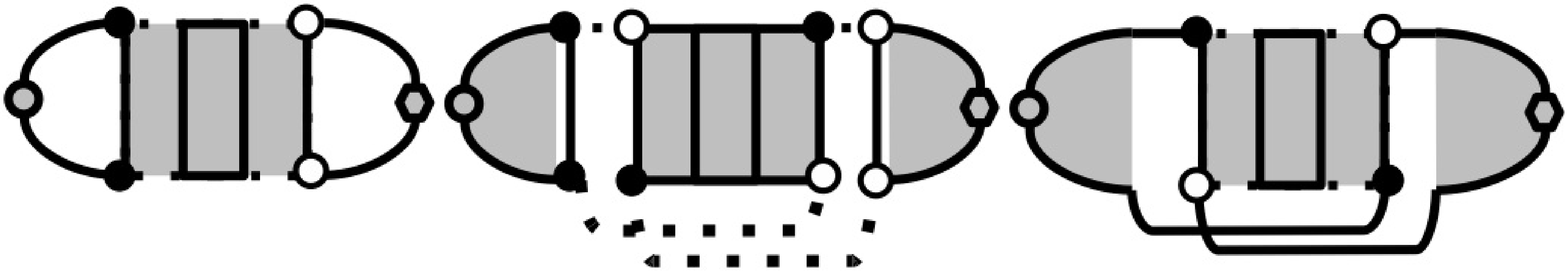}
   \caption{The first diagram is known as diffuson and the two others are known as cooperons.} \label{diagramas.eps}
\end{figure}

We model the system statistical properties using the RMT. Following references \cite{brouwer,nos2}, the resonance $S$-matrix can be parameterized as
\begin{eqnarray}
S(\varepsilon,{\cal B})=TU[1-Q^\dagger R(\varepsilon,{\cal B}) Q U]^{-1}T^\dagger. \label{SMatriz}
\end{eqnarray}
Here, $U$ is the scattering matrix counterpart of an isolated QD, while $R$-matrix (representing the external fields) is the stub counterpart. Eq.(\ref{SMatriz}) covers the total concatenated resonant scattering $S$-matrix process over circular ensembles. To incorporate the symplectic symmetry, we write $ U $ of order $ M $ in the quaternionic form. $M$ stands for the number of resonances of the QD, while $N=N_1+N_2$ is the total number of open channels, also identified with the order of the $S$-matrix. The universal regime requires $M \gg N$. The $Q$-matrix is a projection operador  of order $(M-N) \times M$, while $ T $ of order $N \times M$ describes the channels-ressonances couplings. Their explicit forms read $ Q_{i, j} = \delta_{i +N, j} $ and $ T_{i, j} = \textrm{diag} ( i \delta_{i, j} \sqrt{\Gamma_{1}}, i \delta_{i+N_1, j} \sqrt{\Gamma_{2}})$. The matrix $R$ of order $ (M-N) $ reads $R(\varepsilon,x)=\exp\left(\frac{i E}{M} \sigma^0+\frac{x}{M}X\sigma^0 + i V \right)$ where  $X$ is a Gaussian distributed random matrix, while $E = (\tau_{\rm dwell}/ \hbar) \varepsilon$ and $x^2 = \tau_{\rm dwell}/\tau_{{\cal B}}$ are dimensionless parameters representing $\varepsilon$ and ${\cal B}$. $V$ is a effective deterministic potential that reveals the presence of tunneling probabilities \cite{nos2} on the stub.

Both the dwell time $\tau_{\rm dwell}$ and the magnetic scattering time $\tau_{\cal B}$ are a system specific quantities. $\tau_{\rm dwell}$ is usually expressed in terms of the decay width \cite{FS1} through the Weisskopf argument \cite{Richter}, namely $\gamma = \hbar/\tau_{\rm dwell}= \Delta /(2\pi) \sum_{i=1,2} \sum_{c} \Gamma_{i,c}$, where $\Delta$ is the system mean level spacing and $\Gamma_{i,c}$ is the tunneling probability of the channel $c$ in the lead $i$. In turn, $\tau_{\cal B}^{-1}$ is the rate by which the electron trajectory accumulates magnetic flux in the quantum dot. For chaotic systems, $\tau_{\cal B}^{-1}=\kappa ({\cal A} {\cal B})/\phi_0$, where $\phi_0$ is the unit flux quantum, ${\cal A}$ is the quantum dot lithographic area, and $\kappa$ is a diffusion coefficient that depends on the quantum dot geometry \cite{Pluhar}. This RMT calculation at zero temperature is valid in ballistic chaotic ergodic regime, for which only the irregular borders \cite{SChaos} of the QD contribute to universal fluctuations. This is equivalent to consider only long electronic trajectories, ie, all time scales are much larger than the Ehrenfest time, $\left\{ \tau_{\rm dwell}, \tau_{{\cal B}} \right\} \gg \tau_{\rm Ehrenfert}$.

The matrix $S=\left(\begin{array}{cc}r & t \\t' & r' \\ \end{array}\right)$ can be decomposed in transmission $t(\varepsilon, {\cal B})$ and reflection $r(\varepsilon, {\cal B})$ sub-matrices. In this way, the dimensionless conductance can be written in terms of the $S$-matrix through the Landauer-B\"uttiker formula $g(\varepsilon,x)=\textrm{Tr}(t^\dagger(\varepsilon,x)t(\varepsilon,x))$. Consequently, $g(\varepsilon,x)$ can be formally expanded in powers of $U$. For chaotic QDs in the universal regime, as standard, we assume the matrix elements of $U$ as Gaussian random variables with variance $1/M$. This allows one to express the calculation of moments and cumulants of $g(\varepsilon,x)$ by an integration over the unitary group leading to a diagrammatic expansion in powers of $N^{-1}$ in terms of diffusons (ladder) and cooperons (maximally crossed) diagrams. The method is described for the Dyson ensembles in Ref.\cite{brouwer} and was extended to treat the crossover between symmetry classes \cite{brouwer2,nos2} for the noise, yielding the same results as another approaches \cite{Jacquod}.

Firstly, we reobtain the averaged conductance \cite{nos2} of a chaotic QD. The Fig.[\ref{diagramas.eps}] shows the diagrams contributing to $\left< g(\varepsilon, {\cal B}) \right> $. Their white and black dots represent the indices $i$ and $j$ of the matrix $U$ with entries $U_{ij}$, while the lines represent indices contractions. We obtain $\langle g(x)\rangle = 2\frac{G_1G_2}{G_T}\left(1-\frac{G_1\Gamma_2+G_2\Gamma_1}{G_T G_C}\right)$
where $G_T=G_1+G_2$, $G_C=G_T+2 x^2$ and the $\varepsilon$-dependence disappear in the ensemble average. For pure ensembles, the quantum interference term disappears if $x \rightarrow \infty$ (unitary ensemble) and reaches its maximum value if $x \rightarrow 0$ (orthogonal ensemble). Finite $x$ characterizes the crossover.

The CF $C(\delta \varepsilon, \delta {\cal B})$ is identified with the ensemble averaged covariance $cov\left[g(\varepsilon,x),g(\varepsilon',x')\right]=\left< (g(\varepsilon,x)-\left< g(\varepsilon,x)\right>) (g(\varepsilon',x')-\left< g(\varepsilon',x')\right>)\right>$. We extend the stub method for the new calculation of the general CF, including the quantum interference terms. Using the typical large number of diagrams for CFs \cite{brouwer}, we calculate the averages of four-point scattering matrices, achieving
\begin{widetext}
\begin{eqnarray}
&& \frac{C(\delta\epsilon,\delta X)}{2\frac{G_1^2G_2^2}{G_T^4}}=-\frac{G_1^3\Gamma_2(\Gamma_2-1) +G_2^3\Gamma_1(\Gamma_1-1)}{G_1G_2G_T}\sum_{i=0}^1\frac{1+(2X\delta_{i1}+\delta X)^2}{\left[1+(2X\delta_{i1}+\delta X)^2\right]^2+(\delta \epsilon)^2} + \frac{\left(G_1\Gamma_2 +G_2\Gamma_1 \right)^2}{2G_T^2} \times \nonumber \\
&\times& \sum_{i=0}^1\frac{1}{\left[1+(2X\delta_{i1}+\delta X)^2\right]^2+(\delta \epsilon)^2 }
+ \frac{\left[G_1(\Gamma_2-1) +G_2(\Gamma_1-1) \right]^2}{G_T^2} \sum_{i=0}^1\frac{\left[1+(2X\delta_{i1}+\delta X)^2\right]^2-(\delta \epsilon)^2}{\left\{\left[1+(2X\delta_{i1}+\delta X)^2\right]^2+(\delta \epsilon)^2 \right\}^2}
 \label{general}
\end{eqnarray}
\end{widetext}
where $X=x/2G_T$ and $\epsilon=\varepsilon/\gamma$. Eq.~\eqref{covg} is obtained from the general Eq.~\eqref{general} taking $\{ \Gamma, \delta X\} \to 0$ and ${\cal B}=0$ or ${\cal B} \rightarrow \infty$. Eq.~\eqref{general} without dependence in $\epsilon$, ${\cal B}$, $\delta \epsilon$ and $\delta {\cal B}$ reproduces Ref.\cite{brouwer} for which high tunnel barrier ($\Gamma \ll 1$) doubles the variance. Using Eq.~\eqref{general}, we show that the Onsager-Casimir relations \cite{PB} is not affected by any value of tunneling probabilities, ie, $cov \{[g (\varepsilon,{\cal B})-g (\varepsilon,- {\cal B})]^{2}\} = 0$.

Setting $\delta \varepsilon =0$ and considering pure ensembles, reduces Eq.~\eqref{general} into a one containing only parametric variation with the magnetic field. For simplicity, we choose $\Gamma_{i}=\Gamma$ and $N_{i}=N/2$ and obtain
\begin{eqnarray}
\frac{C(\delta X)}{1/(8\beta)}&=& \frac{2\Gamma\left(1-\Gamma\right)}{1+(\delta X)^2}+\frac{2+\Gamma\left(3\Gamma-4\right)}{\left[1+(\delta X)^2\right]^2}. \label{covgb}
\end{eqnarray}
Eq.~\eqref{covgb} shows clearly that the CF has a square-Lorentzian shape, $C(\delta X) \propto (1+\delta X)^{-2}$, in both ``quasi-closed" and ideal regimes. Furthermore, for any value of $\Gamma$, magnetoconductance CFs in pure ensembles never shows anti-correlation in the semiclassical regime. On the other hand, parametric variations of the energy produces a transition to a highly non-Lorentzian behaviour through the expression
\begin{eqnarray}
\frac{C(\delta\epsilon)}{1/(8\beta)}&=& \frac{3\Gamma\left(2-\Gamma\right)-2}{1+(\delta\epsilon)^2}+\frac{4\left[1+\Gamma\left(\Gamma-2\right)\right]}{\left[1+(\delta\epsilon)^2\right]^2}. \label{covgegamma}
\end{eqnarray}
The Eq.~\eqref{covgegamma} shows that the ideal case, $\Gamma=1$, yields the usual Lorentzian shape. In the quasi-closed regime, it reduces to the anti-correlation of Eq.~\eqref{covg}. For arbitrary $\Gamma$, it leads to the inset of Fig.[1], where one sees a region , in the diagram  $\Gamma \times (\delta \varepsilon/\gamma)$, a normalized anti-correlation with an amplitude of the order $-0.25$ (white region). As the inset indicates, the anti-correlation appears in a large range of tunneling probabilities for sufficiently high parametric variations of electron energy. We point out here that this specific finding is qualitatively similar to the experimental result of Ref.\cite{Zumbuhl} in QD and also to the anti-correlation in the nuclear reaction $p+^{35} Cl \rightarrow \alpha +^{32} S$ (Fig.[6] of Ref.\cite{Alhassid}).
\begin{figure}[h!]
	\centering
	        \vskip-0.4cm
		\includegraphics[width=\columnwidth,height=4.5cm]{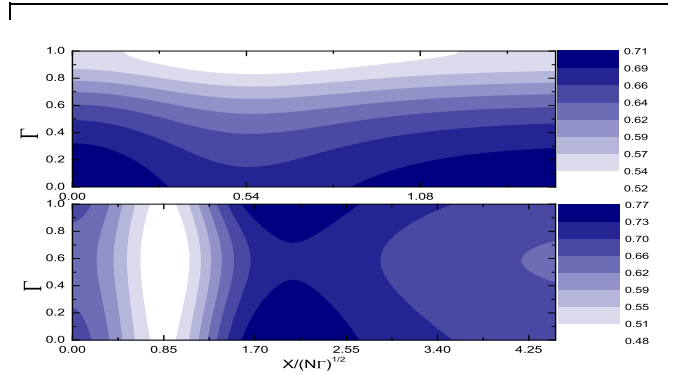}
	\caption{(Color online) Top (bottom) diagram shows the density of peaks $\left< \rho_{\varepsilon}\right>$ ($\left< \rho_{X}\right>$) for parametric variation of the electron energy (perpendicular magnetic field). The darker and lighter regions are explained on the right hand strip.}
	\label{fig:TvsEN=5}
\end{figure}

Support to our analytical findings is provided by numerical simulations employing the Hamiltonian approach to the statistical $S$-matrix \cite{VWZ}, namely
\begin{equation}
\label{eq:SHeidelberg}
S (\varepsilon,{\cal B}) = \mbox{$\openone$} - 2\pi i W^\dagger (\varepsilon - H ({\cal B})+
i \pi W W^\dagger)^{-1} W \;,
\end{equation}
where $H$ is a random Hamiltonian matrix of dimension $M \times M$ that describes the resonant states. In the chaotic regime, for which the number of resonances is very large ($M \rightarrow \infty$), the two scattering matrix formalisms, Eqs.~\eqref{SMatriz} and \eqref{eq:SHeidelberg}, are equivalent \cite{BFB}. We can take $H$ as a member of the Gaussian orthogonal (unitary) ensemble for the symmetric (broken) time-reversal case. The matrix $W$ of dimension $M \times (N_1 + N_2)$ contains the channel-resonance coupling matrix elements. Specifically, for a QD in the chaotic universal regime, the eigenvalues ${\cal W}$ of $WW^{\dagger}$ are connected with $\Gamma$ through the formula \cite{Beenakker} ${\cal W}=M \Delta (2-\Gamma \pm 2 \sqrt{1-\Gamma})/(\pi^2 \Gamma)$. Since the $H$ matrix is statistically invariant under orthogonal ($\beta = 1$) or unitary ($\beta = 2$) transformations, the statistical properties of $S$ depend only on the mean resonance spacing $\Delta$, determined by $H$, and on $W^\dagger W$.

We consider finite $\Gamma_{i,c}$ and separate the matrix $W$ into blocks corresponding to the respective couplings of QD with the two leads, $W=(W_{1},W_{2})$, where $W_{i}$ is a $M \times N_{i}$ matrix. We disregard the direct processes requiring the orthogonality condition $W_{i}^{\dagger} W_{j}= \omega_{i} M \Delta \delta_{ij}/\pi^2$ with $\omega_{i}=\textrm{diag}(\omega_{i,1},\omega_{i,2},\dots,\omega_{i,N_{i}})$. The $\Gamma_{i,c}$ of the channel $c$ and lead $i$ can be written in terms of the diagonal matrix $\omega$ through the relation $\Gamma_{i,c}=\textrm{sech}^{2}(\tau_{i,c}/2)$ with $\tau_{i,c}=-\textrm{ln}(\omega_{i,c})$. We consider equivalent couplings, $\Gamma_{i,c}=\Gamma_{i}$, symmetric contacts, and a chaotic QD. Without loss of generality, we chose the GUE for numerical simulation of the previously mentioned Landauer-B\"uttiker conductance CF, using the transmission sub-matrix of Eq.~\eqref{eq:SHeidelberg}, which can be written as $t(\varepsilon)=-2 \pi i W_{1}^{\dagger}(\varepsilon - H +i \pi W W^\dagger)^{-1} W_{2}$. The numerical simulation (HA) of Fig.[1] shows the conductance autocorrelation function obtained through ${\cal N}_r=2000$ realizations of the unitary $H$ matrices with $M=600$ resonances coupled non-ideally with $60$ open channels.  We also execute another numerical simulation (stub) with the same number of channels, resonances and realizations. All the results coincide nicely as can be seen in the inset of Fig [1]. The Hamiltonian simulation also demonstrates the validity of Weisskopf argument in the ``quasi-closed" limit \cite{Richter} and nicely confirms an auto-correlation length $\gamma=G_{T} \Delta/(2 \pi)$. The discrepancies are very small and are confined within the statistical precision ${\cal N}_r^{-1/2}$.

{\it Alternative Statistical Measures.} -- Eqs.\ \eqref{covg}, \eqref{general}, \eqref{covgb} and \eqref{covgegamma} are the main results of this paper. The statistical sampling required to confirm our predictions for the dimensionless conductance is rather large, making the experimental requirements quite daunting as in \cite{Marcus, Zumbuhl, Alhassid}.  An easier accessible statistical measure has been recently proposed \cite{gabrielprl}: The dimensionless conductance $g$ fluctuates as $\varepsilon$ and $x$ are varied. Let us call the external parameter $z$. Useful statistical information can be extracted from the number of maxima (or minima) of the $g$ in a given interval $[z,z+\delta z]$. Using a scale invariance and maximum entropy principle, we show that the average densities of maxima, $ \langle \rho_ {z} \rangle$ of the fluctuating conductance $g(\varepsilon,x)$ is given by
\begin{equation}
\left< \rho_{z} \right> =\frac{1}{2\pi}\sqrt{-\frac{T_4}{T_2}}; \;\;  T_{j} \equiv \frac{d^j}{d(\delta z)^j} C(\delta z) \bigg|_{\delta z=0}. \label{dens}
\end{equation}

\begin{figure}[h!]
	\centering
	        \vskip-0.4cm
		\includegraphics[width=\columnwidth,height=4.0cm]{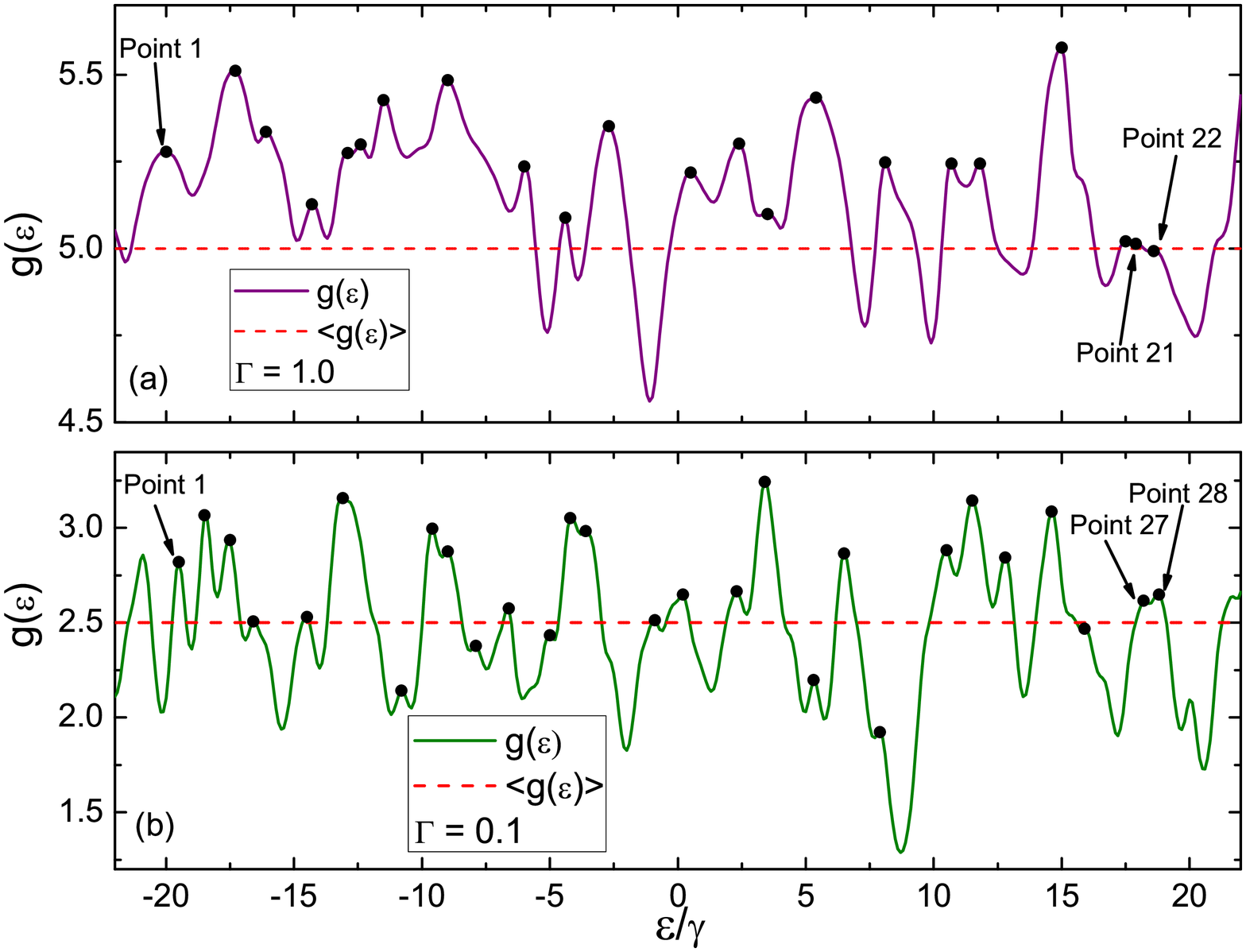}
	\caption{(Color online) Typical dimensionless conductance $g$ as a function of $\varepsilon/\gamma$. Continuous lines for the numerical results for a single realization of $H$. The dots indicate the maxima of $g$ and the dashed line is the $\varepsilon$-independent conductance average.}
	\label{fig:TvsEN=5}
\end{figure}

Firstly, we study electronic energy UCFs traces on pure ensembles, in the presence of finite barriers, using Eqs.~\eqref{covgegamma} and \eqref{dens}. Using an indice $0$ to denote the density of maxima in the ideal case ($\Gamma=1$), we show a relevant transition for which $\left< \rho_{\varepsilon} \right>$ always is larger than $\left< \rho_{\varepsilon}^0 \right> =\sqrt{3}/\pi$, and can be even larger than $\left< \rho_{x}^0 \right> = 3/(\pi \sqrt{2})$. This new result is
\begin{eqnarray}
\langle \rho_\varepsilon \rangle = \frac{\sqrt{3}}{\pi}\sqrt{\frac{9\Gamma^2-18\Gamma+10}{5\Gamma^2-10\Gamma+6}}.
\end{eqnarray}
In $\Gamma \to 0$ limit, the density of maxima significantly increases from $\langle \rho_\varepsilon^0 \rangle=\sqrt{3}/\pi$ to a new chaos number $\langle \rho_\varepsilon \rangle= \sqrt{5}/\pi$, indicating always an amplification (super-density) of the signal of the order of $ 29 \% $, our second main result. We perform a numerical simulation using the Hamiltonian model previously mentioned with a configuration of $60$ open channels and $600$ resonances in an energy range $ \Delta \varepsilon / \gamma \in [-20,20]$. As shown in Fig.[4], the number of maxima, represented by closed points, for the ideal case is $22$ ($\langle \rho_\varepsilon \rangle \approx 0.55$) and the number of maxima for the case $\Gamma=0.1$ is $28$ ($\langle \rho_\varepsilon \rangle \approx 0.71$), confirming nicely our analytical findings.

We identify also a suppression (sub-density) in the density of maxima in the presence of finite tunneling probabilities, but only in the crossover regime. From Eqs.~\eqref{general} and \eqref{dens}, the conductance peaks in the crossover regime as a function of parametric variations of the electron energy (perpendicular magnetotransport field), presents sub-density/super-density transitions, according to the diagram $\Gamma \times X/(N \Gamma)^{1/2}$ upper (lower) of Fig.[3]. In this figure, we display a myriad of values for which the density of maxima can transit as a function of the magnetic field and barriers.

For parametric variation of the perpendicular magnetic field in pure ensembles, we obtain
\begin{eqnarray}
\langle \rho_x \rangle = \frac{\sqrt{3}}{\pi \sqrt{2}}\sqrt{\frac{7\Gamma^2-10\Gamma+6}{2\Gamma^2-3\Gamma+2}} \label{densidademagnetica}
\end{eqnarray}
Eq.~\eqref{densidademagnetica} shows that different intensities of quantum tunneling does not produces large variations in density of maxima. Interestingly, we get $\langle \rho_x \rangle = 3/(\pi \sqrt{2})$ both in the ``quasi-closed" extreme limit, $\Gamma=0$, and in the ideal limit, $\Gamma=1$. According with Fig.[3], large variations of conductance peaks for $\delta X$ occur only in the crossover regime.

{\it Conclusions} - We identify analytically an anti-correlation effect in open chaotic QDs which is valid in a wide range of tunneling barriers. Our result is valid in the universal regime of chaotic ballistic QDs at zero temperatures. The effect modifies the exponential power-spectrum behavior of a Lorentzian \cite{Alhassid}. A different anti-correlation mechanism can also be seen in the power-spectrum of diffusive QDs due to to their typical short trajectories corrections. The anti-correlation that we are reporting occurs strongly in pure Dyson' ensembles for variation of electronic energy, but is appreciably amplified in the presence of magnetoconductance fields. The Weisskopf parameter is still valid even in the ``quasi-closed" regime, leading to a significant amplification of the dwell time of an open chaotic cavity \cite{Hackens}. This deterministic effect induced by barriers changes the poles of the Green functions of the Hamiltonian formalism, leading to an strong amplification of periodicity in a single measure the conductance. Experimental realizations can exhibit $29\%$ of increase in the density of conductance peaks in systems with finite barriers. We also show new universal numbers of quantum chaos. Future investigations using the anti-correlation intrinsic effect and the density of maxima method are nice perspectives to measure the effects of Coulomb blockade due to tunnel barriers \cite{Zumbuhl}, which hinder the electron entrance and exit in the QD. A direct comparison with these experiments at finite temperatures $T$ will be possible considering the typical non-universal (system dependent) Coulomb corrections, proportional to $\Delta$ \cite{bbQD}. These corrections stem from the modification of time scales owing to Coulomb scattering time and can be implemented using RMT \cite{coulomb}.

This work is supported in part by the Brazilian funding agencies CAPES, CNPq, FACEPE, FAPESP and the Instituto Nacional de Ci\^{e}ncia e Tecnologia de Informa\c{c}\~{a}o Qu\^{a}ntica-MCT.

\end{document}